\documentclass[showpacs]{iopart} 
\usepackage[dvips]{graphicx}
\usepackage{cite}
\def\preprint#1{\begin{flushright}#1\end{flushright}}

\newcommand{\nc}{\newcommand}
\def\nn{\nonumber\\}
\def\bea{\begin{eqnarray}}
\def\eea{\end{eqnarray}}
    
\nc{\braket}[1]{\langle\,{#1}\rangle}

\def\wt{\widetilde}

\def\pa{\partial}

\def\Z{\mathbf Z}    
\def\V{{\cal V}}

\begin{document}

\preprint{OIQP-06-13} 

\title{DNA toroid condensation as analytic solutions}

\author{Y Ishimoto$^1$ and N Kikuchi$^2$}
\address{
$^1$ Okayama Institute for Quantum Physics, 1-9-1 Kyouyama, Okayama 700-0015, Japan}
\address{
$^2$ Fachbereich Physik, Martin-Luther-Universit\"at Halle-Wittenberg, D-06099 Halle, Germany
}
\ead{{ishimoto@yukawa.kyoto-u.ac.jp}}
\date{\today}

\begin{abstract}
It now becomes apparent that condensed DNA toroid which emerges in a poor solvent condition can be realised in the framework of the non-linear sigma model on a line segment. In fact, the classical solutions of the model, {\it i.e.}, of the bending potential exhibit toroidal forms, and fit well with the delta-function term of the attractive interaction. The proposed theory is in good agreement with experimental observations that the toroid is the ground state. In this paper, 
we give a rigorous proof that the solutions are indeed the exact solutions of the equations of motion with the first derivatives of the attractive interaction term.
We also show a refined mapping to experimental data, considering the finite size effects of the cross section and of the chain length. 
\end{abstract}


\section{Introduction}

Since in living cells DNA is often packed tightly, 
for instance, inside phage capsids, 
DNA condensation has drawn much attention 
\cite{GS76,B91,B96,YYK98,CVH03,HD01,AB90} 
as well as its biochemical/medical importance in the emerging field of gene  therapy 
\cite{CVH03,HD01}. 
In fact, when we put condensing agents as multivalent cations into DNA
solution, it can cause DNA to undergo the condensation from a
worm-like chain (whip) to toroidal states 
\cite{GS76,B91,B96,YYK98,AB90,CVH03,HD01}. 

A double stranded DNA chain can be modelled by a semiflexible homopolymer chain \cite{DE86,K04,KF72}.
To increase our understanding of such toroid condensation and ``whip-toroid transition'', semiflexible homopolymers in a poor solvent condition (i.e effective interactions between polymer segments are attractive) have been investigated as simple models 
\cite{UO95,MPM04,CW04,NSKY96,NY98,KT99,KTD96,IPB98,GK81,SIGPB03,HDB95,UO96,SMW00,SGM02,PW00,MKPW05}. 
Simulations using Monte Carlo, Langevin approaches or Gaussian variational method, 
calculated phase diagram for the semiflexible chain in a poor solvent 
\cite{MPM04,CW04,NSKY96,NY98,KT99,KTD96,IPB98}.
In most analytic works, existing phenomenological models balance the bending and surface free energies to estimate toroidal properties 
\cite{UO95,GK81,SIGPB03,HDB95,UO96,SMW00,SGM02,PW00,MKPW05}. 
It becomes probable that toroid is the stable lowest energy state --- the ground state. 
We note, however, that the analytic aspects of the works assumed a priori toroidal geometry as the stable lowest energy state with no analytic proof \cite{SMW00}. 
Moreover, compared to the theory of coil-globule transition of flexible chains \cite{LGK78,KF84,DG79,DE86,GK94}, which are well described by field theoretical formalism \cite{LGK78,KF84}, 
there was no simple microscopic theory, which contains the salient physics to demonstrate the whip-toroid transition of the semiflexible polymer.

In our previous work \cite{IK05}, 
we introduced a novel microscopic model for the toroidal condensation and the related conformational transitions of the semiflexible homopolymer chain, using the path integral method in the framework of the non-linear sigma model on a line segment. The proposed theory resolved the difficulties arising from the local inextensibility constraint of the polymer chain \cite{K04,KF72} and from the non-local nature of the attractive interaction along the chain. 
The classical solutions of the non-linear sigma model fit well with the attractive interaction term between polymer segments, and demonstrate the whip-toroid transition of the semiflexible polymer chain. Moreover, predicted mean toroidal radii are in quantitative agreement with experimental data of DNA toroids. However, it has not been proved that the classical solutions are the exact solutions of the equations of motion of the whole system, {\it i.e.}, of the sigma model with the non-local attractive interaction term.

In this paper, reviewing \cite{IK05}, we investigate a single semiflexible homopolymer chain with attraction, using the path integral method and the nonlinear sigma model on the line segment. 
We provide a rigorous proof that the solutions of the model are indeed the exact solutions of the equations of motion of the system, and renew the derivation of the attractive potential energy. 
A new and refined mapping of the mean toroid radii to experimental data of DNA toroid is also given, considering the finite size effects of the cross section and of the chain length.
Note that a detailed discussion on the stability of the solutions is given in \cite{IK05}, so we are not going to discuss this issue in what follows.

\section{Preliminaries}

In the continuum limit, the Green function (end-to-end distribution) of a semiflexible polymer chain
with attractive interactions can be given by the path integral: 
\bea
G(\vec{0},\vec{R};\vec{u}_i,\vec{u}_f;L,W)=
{\cal N}^{-1}\!\!\!\! \int_{\vec{r}(0)=\vec{0}, {\vec{u}(0)=\vec{u}_i}}^{\vec{r}(L)=\vec{R}, {\vec{u}(L)=\vec{u}_f}} 
  \!\!\!\!\!\!\!\!\!\!\!\!\!\!\!\!\!\!\!\!\!\!\!\!\! {\cal D} [\vec{r}(s)]\, e^{- {\cal H}[\vec{r}, \vec{u}, W]} , 
\eea
with a local inextensibility constraint $|\vec{u}(s)|^2 = 1$ 
\cite{K04,KF72}. 
$s$ is the proper time along the polymer chain of total contour length $L$. $\vec{r}(s)$ denotes the pointing vector at the `time' $s$ in our three dimensional space while
$\vec{u}(s)\equiv\frac{\pa \vec{r}(s)}{\pa s}$ corresponds to the unit tangent (or bond) vector at $s$. ${\cal N}$ is the normalisation constant. Following Freed et al. and Kleinert 
\cite{KF84,K04}, the dimensionless Hamiltonian can be
written by ${\cal H}[\vec{r}, \vec{u}, W] =
\int_0^L ds\, \left[ H(s) +
  V_{AT}(s) \right]$ where $H(s)$ and $V_{AT}(s)$ are the local free Hamiltonian and the non-local attractive interaction term, respectively: 
\bea
  H(s) &=& \frac{l}{2} \left\vert \frac{\pa}{\pa s} \vec{u}(s)
  \right\vert^2\!\!\!,\\
  V_{AT}(s) &=& - W \int_0^s d s^\prime \delta \left( \left| \vec{r}(s) 
  - \vec{r}(s^\prime) \right| \right)\!.
  \label{VDWhamiltonian}
\eea
Note that the interaction term is non-local with respect to the proper time $s$ of the path integral.
The persistence length $l$ is assumed to be large enough to realise its stiffness and $W$ is a positive coupling constant
of the attractive interaction between polymer segments. The precise definition of $V_{AT}$ will be given at the beginning of the following section. Thermodynamic $\beta=1/(k_BT)$ is implicitly included in $l$ and $W$. 
In what follows, we express $\vec{r}$ by the unit tangent vector $\vec{u}$ and therefore the Hamiltonian ${\cal H}(\vec{u})$ in terms of $\vec{u}$. 
Hence, the Green function $G\left( \vec{0}, \vec{R};
\vec{u}_i, \vec{u}_f; L, W \right)$ becomes a path integral over $\vec{u}$ with the positive coupling constant $W$, regardless of $\vec{r}$,
\bea
G = \int_{\vec{u}_i}^{\vec{u}_f} {\cal D} [\vec{u}(s)]\, 
    \delta\left({\textstyle \int_0^L ds\, \vec{u}(s) - \vec{R} }\right) \, e^{- {\cal H}[\vec{u}, W]} ,
\eea
where we used $\vec{r}(L)= \int_0^L ds\, \vec{u}(s)$ 
and the Jacobian is absorbed by ${\cal N}$ which is neglected here. 
The delta
function selects out the end-to-end vector.
When $W=0$, our free dimensionless Hamiltonian with the constraint
$|\vec{u}(s)|^2=1$ can be interpreted as the low energy limit of a
linear sigma model on a line segment, or quantum equivalently 
a nonlinear sigma model on a 
line segment, rather than some constrained Hamiltonian system. 


We begin with $O(3)$ nonlinear sigma model on a line segment which is nothing but a quantum mechanics of a limited time $s\in [0,L]$ with the constraint. 
The constraint $|\vec{u}|^2 = 1$ restricts the value of $\vec{u}$ onto a unit sphere $S^2$. Its spherical coordinate decomposition in the free Hamiltonian ${\cal H}(\vec{u})=\frac{l}{2} \int_0^L ds\, \left| \pa \vec{u}(s) \right|^2$ gives the action: 
\bea
 S[\theta_u, \varphi_u] &=& \frac{l}{2} \int_0^L ds\; \left[ (\pa \theta_u)^2 + \sin^2 \theta_u (\pa \varphi_u)^2 \right]. 
\label{nonlinearsigmaaction}
\eea
Minimizing the action in terms of $\theta_u$ and $\varphi_u$ yields the classical equations of motion: 
\bea
 \left[ - \pa^2  + \frac{\sin 2\theta_u}{2 \theta_u} (\pa \varphi_u)^2 \right] \theta_u &=& 0,
 \nn
 \left[ \,\pa^2 + 2 (\pa \theta_u) \cot \theta_u \pa\, \right] \varphi_u &=& 0 .
\label{EOM}
\eea
We then find the following two types of classical solutions: 
\bea
  && \vec{u}(s) = const.
  \nn
  &&or~~
  \nn
  && \theta= \frac{\pi}{2} \quad and \quad \varphi_u = a s + b ,
  \label{classical sol}
\eea
where $a, b$ are constants. 
By symmetry argument, we proved that the above solutions represent the general solutions \cite{IK05}. That is either a constant $\vec{u}(s)$ (rod solution) or a rotation at a constant speed along a great circle (toroidal solution). 
Note that they appear as minimal conformations in the three dimensional $\vec{r}$ space rather than in two dimensions of well-known Euler's elastica.
In what follows, without loss of generality, we set $a\geq 0$ and $b=0$ unless otherwise stated.

Our aim is now to explore classical solutions (\ref{classical sol}) and to study its corresponding conformations in the presence of the attractive interactions. Before we achieve this, we treat more carefully the attractive potential term (\ref{VDWhamiltonian}) in the equations of motion. It might be obvious that the rod solution in (\ref{classical sol}) gives zero to the first derivatives of the attractive potential, because any small change of the solution contributes nothing to the potential. Also, the second solution seems to give zero to the first derivatives of the potential. Because, since its corresponding conformation sits on a single circle, the solution would locally give a bottom of the potential. In the following section, we explicitly prove such zeros of the derivatives in a rigorous way.

\section{The attractive potential part of the equations of motion}

In this section, we show explicitly the first derivatives of the attractive potential which enter the classical equations of motion. We then verify that they vanish in the classical configurations (\ref{classical sol})
and therefore that the configurations are in fact the exact solutions of the equations of motion of the whole system.
The precise definition of the attractive potential term is given in \cite{IK05} by:
\bea
\int_0^L ds\; V_{AT}(s)
 = -W \int_\epsilon^L ds\; \int_0^{s-\epsilon} ds^\prime\; \delta^{re} \left( R(s,s^\prime) \right),
 \label{def:pot}
\eea
where
\bea
R(s,s^\prime)
 \equiv \left| \vec{r}(s) - \vec{r}(s^\prime) \right| 
 = \left| \int_{s^\prime}^{s} dt\; \vec{u}(t) \right|.
\label{R}
\eea
$\epsilon$ is the ultraviolet (short range) cut-off and $\delta^{re}$ is the renormalised delta function which matches Kronecker's delta in its discrete correspondence. In our case, it is 
\bea
\delta^{re} \left( R(s,s^\prime) \right)
=
 R' (s,s^\prime)\; \delta \left( R(s,s^\prime) \right),
\label{re delta}
\eea
where $R'=\frac{d}{ds^\prime} R$.

The first $\theta$ derivative of the attractive potential can be given by:
\bea
&&\hspace{-20pt}
\frac{\delta}{\delta \theta(s_0)} \int_\epsilon^L ds\; \int_0^{s-\epsilon} ds^\prime\; \delta^{re} \left( R(s,s^\prime) \right)
\nn
 &=& \int_\epsilon^L ds\; H_{step}(s-s_0) 
   \int_0^{s-\epsilon} ds^\prime\; H_{step}(s_0-s^\prime)
   \frac{\delta}{\delta \theta(s_0)}  \delta^{re} \left( R(s,s^\prime) \right)
\nn
 &=& \int_{max(s_0,\epsilon)}^L ds\; 
   \int_0^{min(s_0,s-\epsilon)} ds^\prime\; 
   \frac{\delta}{\delta \theta(s_0)}  \delta^{re} \left( R(s,s^\prime) \right) ,
   \label{theta-D}
\eea
where $H_{step}(s)$ is the Heaviside step function. 
The first derivative of the renormalised delta function (\ref{re delta}) can be further reduced to:
\bea
\frac{\delta}{\delta \theta(s_0)}  \delta^{re} \left( R(s,s^\prime) \right)
  &=& \frac{d}{d s^\prime}\left[
  \delta \left( R(s,s^\prime) \right) \left( \frac{\delta R(s,s^\prime)}{\delta \theta(s_0)} \right) \right] .\label{theta_1stderiv001}
\eea
Substituting (\ref{theta_1stderiv001}) into (\ref{theta-D}), one obtains
\bea
&&\hspace{-20pt}
\frac{\delta}{\delta \theta(s_0)} \int_0^L ds\; V_{AT}(s)
\nn
   &=& \int_{min(s_0+\epsilon,L)}^L ds\; \Biggl\{
  \delta \left( R(s,s_0) \right) \left( \frac{\delta R(s,s_0)}{\delta \theta(s_0)} \right) \Biggr\} 
   \nn&&
    - \int_{max(s_0,\epsilon)}^L ds\; \left\{ \delta \left( R(s,0) \right) \left( \frac{\delta R(s,0)}{\delta \theta(s_0)} \right)  \right\}
  \nn&&
   + \int_{max(s_0,\epsilon)}^{min(s_0+\epsilon,L)} ds\; \Biggl\{
  \delta \left( R(s,s-\epsilon) \right) \left( \frac{\delta R(s,s-\epsilon)}{\delta \theta(s_0)} \right) \Biggr\} .
  \label{theta der}
\eea
Similarly, the first $\varphi$ derivative of the potential can be given by replacing all the $\theta(s_0)$ derivatives with those of $\varphi(s_0)$.
It should be stressed here that when the chain has no interactions between segments, {\it i.e.}, $\delta(R)=0$, all the above terms trivially vanish.
Therefore, the equations of motion are given by eq.(\ref{EOM}).
For the detailed derivation of eqs.(\ref{theta_1stderiv001}, \ref{theta der}), and  another expression of (\ref{theta der}), refer to \ref{AppA}.

First, let us calculate the first $\theta$ derivative of the attractive potential in the classical configurations.
A naive calculation of the first derivative of $R$ shows that it is, in general, of the form $1/R$ and becomes singular for nonzero values of $\delta(R)$.
Therefore, it sometimes fails to estimate exactly the first derivative of the potential. This is indeed the case for the $\theta$ derivative. Instead, we take rather an intuitive way for the exact estimate of the $\theta$ derivative.

Around the classical configurations, $\theta$ is always around $\pi/2$, thus the small $\theta$ fluctuation at $s_0$ always results in an infinitesimal shift of $R(s_1,s_2)$ in the $\theta$ direction for $s_1 \geq s_0 \geq s_2$. Especially when $R(s_1, s_2)=0$, the fluctuation can be expressed by the sinusoidal perturbation
\bea
   \delta R(s_1,s_2) = \left| \sin\left(\theta(s_0) - \frac{\pi}{2}\right) \right|, \label{delR}
\eea
which clearly shows that the function $R$ is not smooth at $R=0$. Substituting $\theta(s_0) = \pi/2 + \delta\theta(s_0)$ into (\ref{delR}),
we have
\bea
\delta R(s_1,s_2) = \left| \sin(\delta\theta(s_0)) \right| = \left| \delta\theta(s_0) \right| .
\eea
Therefore,
\bea
\frac{\delta R(s_1,s_2)}{\delta \theta(s_0)} = sgn\left( \theta(s_0) - \theta_{classical}(s_0) \right),
\eea
where $\theta_{classical}(s_0) = \pi/2$. By the standard definition of the $sgn$ function, $sgn(0)=0$, it follows that
\bea
\frac{\delta R(s_1,s_2)}{\delta \theta(s_0)} \Biggr\vert_{\theta(s_0)=\frac{\pi}{2}} = 0 .
\eea
Hence, the first $\theta$ derivative vanishes in the classical configurations:
\bea
 \frac{\delta}{\delta \theta(s_0)} \int_0^L ds V_{AT}(s) \biggr\vert_{classical} 
  = 0 . 
\eea

An infinitesimal shift in the $\varphi$ direction to $R$ is not as simple as the one in the $\theta$ direction. Rather, a straightforward calculation shows that the first $\varphi$ derivative of the potential vanishes as well in the classical configurations. 
By substituting (\ref{classical sol}) into (\ref{R}), the function $R$ can be explicitly written down for the classical solutions (\ref{classical sol}):
\bea
  R(s,s^\prime) \bigr\vert_{classical}
   = \frac{2}{a} \left|\,\sin \frac{a}{2}(s-s^\prime) \right| .
\label{Rcl}
\eea
Similarly, its $\varphi$ derivative is given by:
\bea
  \frac{\delta R(s,s^\prime)}{\delta \varphi(s_0)} \biggr\vert_{classical}
  \hspace{-10pt}
  =  - \frac{\cos a(s-s_0) - \cos a(s^\prime - s_0)}{2 \left|\,\sin \frac{a}{2}(s-s^\prime) \right|} . \label{First_phi_derv001}
\eea
When $R$ approaches zero, that is, $s = s^\prime + \frac{2\pi m}{a} + \varepsilon$ with $|\varepsilon| \ll 1$ and $m\in\Z_+$, then the l.h.s. of eq.(\ref{First_phi_derv001}) becomes 
\bea
l.h.s.
  &=& - \frac{\cos a(s^\prime + \varepsilon - s_0) - \cos a(s^\prime-s_0)}{2 \left|\,\sin \frac{a \varepsilon}{2} \right|}
\nn
  &=& \left| \sin \frac{a \varepsilon}{2} \right|
      \cos a(s^\prime\! -\! s_0)
  + sgn\left( \sin \frac{a \varepsilon}{2} \right)
    \cos \frac{a \varepsilon}{2}
    \sin a(s^\prime\! -\! s_0) .
\eea
Therefore, the first $\varphi$ derivative turns out to be well-defined.
For the specific values of $s^\prime$, one finds 
\bea
  \frac{\delta R(s,s_0)}{\delta \varphi(s_0)} 
  &=& \left| \sin \frac{a \varepsilon}{2} \right| , 
\nn
  \frac{\delta R(s,0)}{\delta \varphi(s_0)} 
  &=& \left| \sin \frac{a \varepsilon}{2} \right| \cos (a s_0)
  - sgn\left( \sin \frac{a \varepsilon}{2} \right)
    \cos \frac{a \varepsilon}{2}
    \sin (a s_0) .
\label{specific derivatives}
\eea
Now we are ready to calculate the derivative around a singularity of $\delta(R(s,s_0))$:
\bea
\int_{-\delta}^{\delta} d\varepsilon\; \delta\left( \frac{2}{a}\left| \sin \frac{a \varepsilon}{2} \right| \right) \left| \sin \frac{a \varepsilon}{2} \right|
 &=& \int_{-\delta'}^{\delta'} dX\; \delta(X) \frac{ \frac{a}{2} \left| X\right|}{\sqrt{1- \frac{a^2 X^2}{4}}} 
 \nn
 &=& 0 , 
\eea
where $X=\frac{2}{a} \sin(a\varepsilon/2)$, $\delta$ and $\delta'$ pick up the neighbourhood of the zero only ({\it i.e.}, $0< \delta \ll 1$), and the identity $\delta(|X|)= \delta(X)$ is used. 
The same holds for the first term of $R(s,0)$ in (\ref{specific derivatives}).
As for the second term, one should be aware from the existence of the $sgn$ function that it behaves odd around the zeros of the delta function. Accordingly, this term does not contribute to the final result. Finally, $R(s,s-\epsilon) \bigr|_{classical} = \epsilon >0$ and $\delta(\epsilon)=0$ so that the third term in (\ref{theta der}) vanishes.
Hence, the first $\varphi$ derivative of the attractive potential also vanishes: 
\bea
 \frac{\delta}{\delta \varphi(s_0)} \int_0^L ds V_{AT}(s) \Biggr\vert_{classical} 
  = 0 . 
\eea


\section{Whip, toroid, and the potential energy}
\label{sec:potential}

Now let us consider the attractive potential (\ref{VDWhamiltonian}) itself. 
It is difficult to interpret it in the context of quantum theory due to its non-local nature along the polymer chain. 
However, we can derive its explicit form in the classical configurations (\ref{classical sol}). Substituting (\ref{Rcl}) into (\ref{def:pot}), one finds
\bea
   \int_0^L ds\; V_{AT}(s)
   &=& -W \int_{\epsilon}^{L} ds\; \int_{0}^{s-\epsilon} ds^\prime\; \delta^{re} \left( \left| \sin \frac{a}{2}(s-s^\prime) \right| \right)
\nn
   &=& -W \int_{\epsilon}^{L} ds\; \int_{0}^{s-\epsilon} ds^\prime\; \delta^{re} \left( (s-s^\prime) {\rm ~mod~} \frac{2\pi}{a} \right)
   \nn
   &=& -W \int_{\epsilon}^{L} ds\; \left[ \frac{as}{2\pi} \right]
   \nn
   \nn
   &=&  - W \left[ \frac{2\pi}{a} \sum_{k=1}^{N(L)-1} k + \frac{2\pi}{a} \left( \frac{aL}{2\pi} - N(L) \right) N(L) \right] 
   \nn
   &=&  \frac{\pi W}{a} N(L) \left(N(L)+1\right) - W L\, N(L)
   , 
   \label{Hamilwithattraction}
\eea
where $\delta^{re} ( const \times x)=\delta^{re}(x)$ is used, and $N(s) \equiv [as/2\pi]$ by Gauss' symbol\footnote{
Gauss' symbol $[x]$ gives the greatest integer that is not exceeding $x$.
}. 
$a\epsilon/2\pi \ll 1$ is assumed without loss of generality. 
Note that $N(L)$ represents the winding number of the classical solution (\ref{classical sol}) along a great circle of $S^2$. 
Therefore, the form of the potential can be thought of as `topological' terms from the winding number.

The non-zero winding number of the classical solution in the $\vec{u}$
space means that the polymer chain winds in the $\vec{r}$ space as
well. That is, when $a>\frac{2\pi}{L}$, configurations around the
second classical solution (\ref{classical sol}) start forming a toroidal shape since
\bea
 \vec{r}(s) = \left( \begin{array}{c} 
    \frac{1}{a} \left\{ \sin (as +b)-\sin (b) \right\} \\
    -\frac{1}{a} \left\{ \cos (as +b)-\cos (b) \right\} \\
    const.
  \end{array} \right) ,
\eea
and stabilise itself by attracting neighbouring segments. We call such classical solutions the ``toroid states.''
Whenever $a$ increases and passes through the point $\frac{2\pi n}{L}$ for $n\in\Z_+$, another toroid state appears with the increased winding number $n$. Note that the radius of the toroid state is given by $\frac{1}{a}$. 
When the chain of contour length $L$ winds $N(L)$ times we have the $N(L)$ circles of each length $\frac{2\pi}{a}$ and the rest $\left( L-\frac{2\pi}{a}N(L) \right)$. 
The first and second terms in eq.(\ref{Hamilwithattraction}) result from the former and the latter, respectively. 
When $0<a\leq \frac{2\pi}{L}$, the chain cannot wind like the toroid states. Both ends of the chain are not connected to each other, thus can move freely as well as any other parts of the chain fluctuate. 
As long as the total energy of the chain does not exceed the bending energy of $\frac{2\pi^2 l}{L}$ at {}$a=\frac{2\pi}{L}$, 
they can whip with zero winding number. 
We call such low-energy states the ``whip states.'' 
Although the definition includes fluctuations around the classical solutions, unless otherwise stated, 
we primarily refer to the classical solutions of such states, which are rather bowstrings than whips.
\begin{figure}[h]
\centering
\includegraphics[width=8cm]{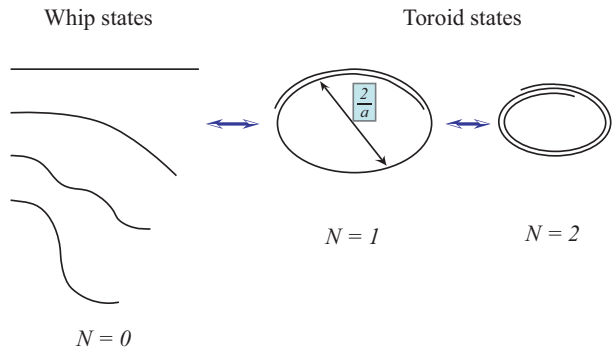}
 \caption{
 The whip ($N=0$) and toroid states ($N\geq 1$).
}
      \label{fig:toroid-whip}
\end{figure}

\section{Energy levels of the analytic solutions}

We now explore the exact energy levels of the whip and toroid states, and
briefly mention the conformational transitions between these states. The dimensionless Hamiltonian of the classical solutions (\ref{classical sol}) is a function of $l, L, W$ and $a$: 
\bea
{\cal H}_{cl}
(a,l,L,W)
\equiv 
\frac{Ll}{2} a^2\!+\!\frac{\pi W }{a} N(L) \left(N(L)+1\right)\!-\!W L\, N(L) , 
\label{Hamiltoniancl}
\eea
with the definition of $\frac{N(L)}{a}=0$ for $a=0$.
It is convenient to rewrite this in terms of the conformation parameter $c \equiv \frac{W}{2l}\left( \frac{L}{2\pi} \right)^2$ and $x \equiv \frac{aL}{2\pi}$, and thus with $[x] = N(L)$:
\bea
  {\cal H}_{cl}(a,l,L,W) &=& WL\,{\cal H}(c,x) 
  = \frac{2\pi^2 l}{L} \left( 4 c\, {\cal H}(c,x) \right),
\eea
where 
\bea
  {\cal H}(c,x)
  &\equiv& \frac{x^2}{4c} + \frac{[x] \left( [x]+1 \right)}{2 x} -  [x].
\eea
Therefore, the function ${\cal H}(c,x)$ can solely express the shape of the energy levels.
Without loss of generality, we assume $x\geq 0$. 

\begin{figure}[h]
\centering
\includegraphics[width=8.0cm]{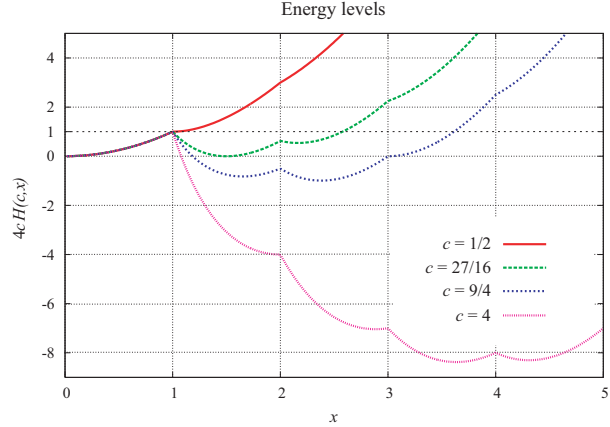} 
 \caption{The dependence of the energy $4c {\cal H}(c,x)$ on $x=a L/2\pi$ and $c$. The unit in energy is set to be the maximum bending energy of the whip state.}
      \label{fig:energy level}
\end{figure}

Let us consider a case with $L$, $W$, and $l$ fixed.
By definition, 
${\cal H}(c,x)$ is continuous in the entire region of $x\geq 0$ and is smooth in each segment: $x \in [ n, n+1 ]$ for $n\in \Z_{\geq 0}$.
However, it is not smooth at each joint of the segments: $x \in \Z_+$. We
plot, in Fig.\ref{fig:energy level}, the energy levels as a function of
$x$ for different values of $c$, showing qualitative agreement with Conwell et al. for DNA condensation \cite{CVH03}.
Given $[x]=N(L)$ is fixed, the Hamiltonian (\ref{Hamiltoniancl}) takes a
minimum at $x=x_c(N(L))\equiv\left( c N(L)(N(L)+1) \right)^{1/3}$.
When $ N(L) <x_c(N(L))< N(L)+1$, ${\cal H}(c,x)$ behaves quadratic in $x$ and $x=x_c(N(L))$ indeed gives its minimum in the $N(L)$th segment. 
In other words, the condition of a (meta-) stable state in the $N(L)$th segment turns out to be 
\bea
N_L(c) < &N(L)& < N_U(c)  \quad {\rm for~} c\geq 4,
\nn 
1\leq &N(L)& < N_U(c)  \quad {\rm for~} 0\leq c < 4,
\eea
where $N_L(c) \equiv \frac{c}{2} \left( 1 - \frac{2}{c} +
\sqrt{1-\frac{4}{c}} \right)$, and $N_U(c) \equiv \frac{c}{2} \left( 1
+ \sqrt{1+\frac{4}{c}} \right)$. 

When $N_U(c) \leq 1$ ({\it i.e.}, $c \leq \frac12$), 
the above second condition vanishes
and thus the whip states only survive at low energy. 
Therefore, at the critical value of $c=\frac12$, the whip phase to
whip-toroid co-existence phase transition may occur. 
When $c > \frac12$, there always exists at least one (meta-) stable state. 
The number of minima are roughly given by the width of the region for $N$, {\it i.e.}, $N_U(c) - N_L(c)$. 
For example, when $c > 9/4$, either $N_U(c) - N_L(c)>3$ or $N_U(c)>3$ with the lower bound being 1, 
hence there are at least three minima with different positive winding numbers. 
This occurrence of the multiple local minima and stable states is because the energy is given by the balance between the bending energy and the attractive potential energy. The former is a monotonically increasing function of $x$ while the latter is monotonically decreasing but not smooth function of $x$. 
Note that 
the number of minima may be reduced in some cases, 
for example, 
with some finite size effect on.
The existence of multiple minima indicates the first order transitions between these stable states. 
In fact, 
when the energy of the $N=1$ stable toroid state becomes zero, 
the whip dominant to toroid dominant phase transition occurs. 
Such a value of $c$ is $27/16$. 
When $c>27/16$, the toroid states will dominate the action. 
Further discussions on the transitions can be seen by changing the value of $c$ and are the subject of a forthcoming publication.

Similarly, one can deduce the inequality for the critical values of $c$:
\bea
c_L^{(N)} < c < c_U^{(N)} , \label{critical_c_inequality}
\eea 
where
$c_L^{(N)} \!\!=\! \frac{N^2}{N+1}$ and 
$c_U^{(N)} \!\!=\! \frac{(N+1)^2}{N}$.
In this region, the $N$th segment has a minimal and (meta-) stable point. 
From this, one can approximate the relation between the conformation parameter $c$ and the dominant toroid winding number $N_c$. 
Furthermore, it leads to the asymptotic relation for either large $c$ or large $N_c$: 
\bea
   c &\simeq& \frac{c_U^{(N_c)} + c_L^{(N_c)}}{2}
   = N_c + \frac12 + O\left(\frac{1}{N_c}\right) \simeq N_c.
\eea
The first $(\simeq)$ is the approximation and the second one denotes the asymptotic relation.

\section{Comparison to experiments and concluding remarks}

For large $c$, the ground state --- the dominant toroid state of the winding number $N_c$ can be estimated by the inequality relation (\ref{critical_c_inequality}) of $c$. It reads $N_c \simeq c$. 
Using this, we can estimate the radius of our ideal toroid: $r_c = \frac{L}{2\pi N_c} = \frac{4\pi l}{W L}$. Our ideal toroids have zero thickness, 
but the finite size effect of their cross sections can be approximated 
by the hexagonally arranged DNA chains with a van der Waals type interaction, {\it i.e.}, with the effective nearest neighbour interactions. 
In the case of $N(L)\geq 4$,
it leads to the modified Hamiltonian \cite{IK05}:
\bea
{\cal H}_{cl} (a,l,L,W)
 &=& \frac{Ll}{2}a^2 - \frac{2\pi W}{a} \V(N(L))
\nn&&
- \frac{2\pi W}{a} \! \left(\frac{aL}{2\pi} -N(L) \right) Gap(N(L)), 
\label{modH}
\eea
where $\V (N) \equiv 3 N -2 \sqrt{3} \sqrt{N-1/4}$ and $Gap(N)\equiv \V (N+1)- \V (N)$.
Following the same procedure for $c_L^{(N)}< c <c_U^{(N)}$, we find 
$
 N_c \simeq \left( 2\sqrt{3} c \right)^{\frac{2}{5}} 
\label{N_c hex}
$
for large $c$. 
By defining the mean toroid radius ({\it i.e.}, the average of inner and outer radii) by $r_c\equiv \frac{L}{2\pi N_c}$, we now estimate the mean radius of the toroid in a physical system in detail. 
A coupling constant can be given by
$W\!=\!\frac{1}{l_m}\!\!\left(\frac{k\epsilon_{w}}{k_BT}\right)$ where $k$
is the number of the electric dipoles in a monomer segment, which create van der Waals interactions of the magnitude $\epsilon_{w}$. $l_m$ denotes the length of the monomer along the chain contour,
taken to be $l_m\simeq5\,bp=1.66\,nm$ in the end.
Substituting $N_c$ and the above, we obtain the asymptotic relation up to numerical factor:
\bea
r_c \simeq 
 {\left(6\pi\right)}^{-\frac15}L^{\frac15}
  {\left(\frac{l}{W}\right)}^{\!\!\frac25}
 ={\left(6\pi\right)}^{-\frac15}L^{\frac15}
  {\left(l_ml\right)}^{\frac25}
  {\left(\frac{k\epsilon_{w}}{k_BT}\right)}^{\!\!-\frac25}.
\label{asym}
\eea
This $L^{\frac15}$ dependence agrees with the proposed exponent in the
asymptotic limit \cite{SIGPB03,SGM02,MKPW05,C05,UO95}.

We estimate the mean toroidal radius of T$4$ DNA in low ionic
conditions reported in \cite{YYK98}. 
Substituting $L=56.44{\mu}m$, $l\,{\simeq}\,50\!\sim\!60\,nm$, and $l_m$ into (\ref{asym}), the mean radius of the toroid is found to be 
$
r_c = 29.03 B^{-\frac25} \sim 31.23 B^{-\frac25} [nm],
$
where $B\equiv \frac{k\epsilon_{w}}{k_BT}$. 
This result is in good agreement with the experiment $r_c \simeq 28.5$ $nm$ for $B \,{\approx}\, 1.15$. 
The same argument for Sperm DNA packaged by protamines ($L=20.4{\mu}m$) \cite{B96}
gives the analytic value 
$
r_c = 23.69 B^{-\frac25} \sim 25.48 B^{-\frac25} [nm],
$
which also agrees with an experimental result $r_c{\simeq}26.25$ $nm$ for $B \,{\approx}\, 0.851$. 
Note that the latter has a larger diameter of the segment and is expected to have the weaker interaction with smaller $B \,{\approx}\, 0.851 $.

\begin{figure}[h]
\centering
\includegraphics[width=8.0cm]{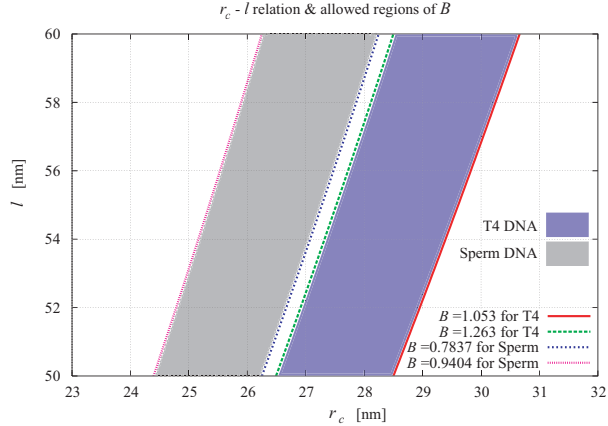} 
 \caption{The relations between $r_c$ and $l$ for T4 DNA with $B=1.053$ and $B=1.263$, and for Sperm DNA with $B=0.7837$ and $B=0.9404$. The two regions are given when $B$ is changed continuously from one to the other.
}
      \label{fig:finite r}
\end{figure}

The above mappings are given on the basis of the asymptotic relation (\ref{asym}), but we can find the range of $B$ numerically without using the asymptotics. The relation $r_c = L/(2\pi N_c)$ means $N_c$ is given by the inverse of $r_c$ up to constant for a fixed contour length $L$.
So, the approximation, $c\simeq \left( c_U^{(N_c)} + c_L^{(N_c)}\right)/2$, leads to $l/B$ as a function of $r_c$: 
\bea
  \frac{l}{B} 
  &=& \left[ \frac{\sqrt{3} L^2}{\pi^2 l_m} \right]
  \frac{ {\wt r_c}^{3/2} \left\{ (1+ {\wt r_c} ) \sqrt{1- {\wt r_c}/4}
  - \sqrt{1+ 3{\wt r_c}/4} \right\} }
  { 1 + (1 + {\wt r_c})^3 } , 
\eea
where ${\wt r_c} \equiv 2\pi r_c/L$.
Therefore, we can plot the $r_c$ - $l$ relations for the fixed values of $B$ and $L$ (Fig.\ref{fig:finite r}).
It reads $B= 1.05 \sim 1.26 \approx 1.16$ ($B_{asympt} \,{\approx}\, 1.15$) for the T4 DNA and $B=0.784 \sim 0.940 \approx 0.862$ ($B_{asympt} \,{\approx}\, 0.851$) for the Sperm DNA, respectively. 

The exponents $\nu$ predicted in the literature for $r_c \sim L^{\nu}$ are $\nu=\frac15$ in most cases \cite{SIGPB03,SGM02,MKPW05,C05,UO95}. 
However, they are inconsistent with the experimentally well known observation that the radius is independent of the chain length \cite{B91,B96,YYK98,AB90}.
This might suggest that the real interaction is not van der
Waals like, or at least is not a single van der Waals type interaction.
It should be noted here that combinations of our ideal toroid and its finite size effect can give a range of $\nu = -1 \sim \frac15$ in some region.
An interesting remark is that when we apply Coulomb like interactions to our approximation, we observe that the radius nearly remains constant as $L$ changes. The precise analysis is to be studied and given elsewhere.


So far we have dealt with the classical solutions, which are derived
from the first derivative of the action. Thus, they may correspond
to the global/local minima of the action in the configuration space. 
In fact, when $c\geq 4$, the toroid states become stable under the fluctuations,
since deviations from such toroid states will cost a large penalty in energy. Its detailed discussion is given in \cite{IK05}.
Note however that, when $c$ is small, 
a number of the whip states become equally or more probable than any toroid state. 
Another interesting question will be how the derivatives of the potential  behave when the configuration be deviated a little, but not infinitesimally, from the classical configurations. 
To answer this question is 
beyond the scope of this paper.

\vspace*{10pt}
\noindent
{\bf Acknowledgments}

{The authors are grateful to K. Binder and his group for their stimulating discussions and hospitality in Mainz. One of the authors (Y.I.) is grateful to K. Nagayama for his discussions and encouragement. The other (N.K.) is grateful to S. Stepanow and S. Trimper for their stimulating discussions and hospitality in Halle, and acknowledges the Deutsche Forschungsgemeinschaft for financial support.}

\appendix
\section{Relevant calculations for the first derivatives of the potential}
\label{AppA}

The attractive potential energy is symmetric under the exchange of $s \leftrightarrow s^\prime$, since $R(s,s^\prime)=R(s^\prime,s)$. Therefore, one can rewrite or regard the double integral in (\ref{def:pot}) as
\bea
&&\hspace{-20pt}
  \int_\epsilon^L ds\; \int_0^{s-\epsilon} ds^\prime\;
=  \frac12 \left(
  \int_\epsilon^L ds\; \int_0^{s-\epsilon} ds^\prime\;
  + \int_0^{L-\epsilon} ds\; \int_{s+\epsilon}^L ds^\prime\;
  \right) . 
\label{double}
\eea
On the other hand,
the first derivative of the renormalised delta function can be reduced:
\bea
&&\hspace{-20pt}
\frac{\delta}{\delta \theta(s_0)}  \delta^{re} \left( R(s,s^\prime) \right)
\nn
  &=& \frac{\delta R'(s,s^\prime)}{\delta \theta(s_0)} \delta \left( R(s,s^\prime) \right)
  + R'(s,s^\prime) \frac{\delta}{\delta \theta(s_0)} \delta \left( R(s,s^\prime) \right)
\nn
  &=& \frac{d}{d s^\prime} \left( \frac{\delta R(s,s^\prime)}{\delta \theta(s_0)} \right)
   \delta \left( R(s,s^\prime) \right)
  + \frac{d R(s,s^\prime)}{d s^\prime} \; \delta^{(1)} \left( R(s,s^\prime) \right)
    \frac{\delta R(s,s^\prime)}{\delta \theta(s_0)} 
\nn
  &=& \Biggl[
  \delta \left( R(s,s^\prime) \right) \frac{d}{d s^\prime}\!\left( \frac{\delta R(s,s^\prime)}{\delta \theta(s_0)} \right)
  + \frac{d}{d s^\prime} \Bigl\{ \delta \left( R(s,s^\prime) \right) \Bigr\}
    \frac{\delta R(s,s^\prime)}{\delta \theta(s_0)} \Biggr]
\nn
  &=& \frac{d}{d s^\prime}\left[
  \delta \left( R(s,s^\prime) \right) \left( \frac{\delta R(s,s^\prime)}{\delta \theta(s_0)} \right) \right] , 
\eea
where $\delta^{(1)}(x)$ is $d \delta(x)/dx$, and the $s^\prime$ and $\theta$ derivatives are assumed to be independent.
Substituting this into eq.(\ref{theta-D}), one obtains
\bea
&&\hspace{-20pt}
\frac{\delta}{\delta \theta(s_0)} \int_0^L ds\; V_{AT}(s)
\nn
   &=& \int_{min(s_0+\epsilon,L)}^L ds\; \Biggl\{
  \delta \left( R(s,s_0) \right) \left( \frac{\delta R(s,s_0)}{\delta \theta(s_0)} \right) \Biggr\} 
  \nn&&
   + \int_{max(s_0,\epsilon)}^{min(s_0+\epsilon,L)} ds\; \Biggl\{
  \delta \left( R(s,s-\epsilon) \right) \left( \frac{\delta R(s,s-\epsilon)}{\delta \theta(s_0)} \right) \Biggr\} 
   \nn&&
    - \int_{max(s_0,\epsilon)}^L ds\; \left\{ \delta \left( R(s,0) \right) \left( \frac{\delta R(s,0)}{\delta \theta(s_0)} \right)  \right\}.
\eea
Here, 
if we revive the relation (\ref{double}), we find the symmetric expression for the first derivative of the potential:
\bea
&&\hspace{-20pt}
\frac{\delta}{\delta \theta(s_0)} \int_0^L ds\; V_{AT}(s)
\nn
   &=& - \int_0^{max(s_0-\epsilon,0)} \hspace{-15pt} ds\; 
   \Biggl\{
  \delta \left( R(s,s_0) \right) \left( \frac{\delta R(s,s_0)}{\delta \theta(s_0)} \right) \Biggr\} 
   + \int_{max(s_0+\epsilon,L)}^L \hspace{-15pt} ds\; 
   \Biggl\{
  \delta \left( R(s_0,s) \right) \left( \frac{\delta R(s_0,s)}{\delta \theta(s_0)} \right) \Biggr\} 
\nn&&
    - \int_{max(s_0,\epsilon)}^L \hspace{-15pt} ds\; \left\{ \delta \left( R(s,0) \right) \left( \frac{\delta R(s,0)}{\delta \theta(s_0)} \right)  \right\}
    + \int_{0}^{min(L-\epsilon,s_0)} \hspace{-15pt} ds\; \left\{ \delta \left( R(L,s) \right) \left( \frac{\delta R(L,s)}{\delta \theta(s_0)} \right)  \right\} \!\!.
\eea
Similarly, the first $\varphi$ derivative of the potential is given by
\bea
&&\hspace{-20pt}
\frac{\delta}{\delta \varphi(s_0)} \int_0^L ds\; V_{AT}(s)
\nn
   &=& - \int_0^{max(s_0-\epsilon,0)} \hspace{-15pt} ds\; 
   \Biggl\{
  \delta \left( R(s,s_0) \right) \left( \frac{\delta R(s,s_0)}{\delta \varphi(s_0)} \right) \Biggr\} 
   + \int_{max(s_0+\epsilon,L)}^L \hspace{-15pt} ds\; 
   \Biggl\{
  \delta \left( R(s_0,s) \right) \left( \frac{\delta R(s_0,s)}{\delta \varphi(s_0)} \right) \Biggr\} 
\nn&&
    - \int_{max(s_0,\epsilon)}^L \hspace{-15pt} ds\; \left\{ \delta \left( R(s,0) \right) \left( \frac{\delta R(s,0)}{\delta \varphi(s_0)} \right)  \right\}
    + \int_{0}^{min(L-\epsilon,s_0)} \hspace{-15pt} ds\; \left\{ \delta \left( R(L,s) \right) \left( \frac{\delta R(L,s)}{\delta \varphi(s_0)} \right)  \right\}\!\!.
\eea



\vspace*{10pt}
\noindent
{\bf References}
\vspace*{3pt}

\begin{thebibliography}{32}
\expandafter\ifx\csname natexlab\endcsname\relax\def\natexlab#1{#1}\fi
\expandafter\ifx\csname bibnamefont\endcsname\relax
  \def\bibnamefont#1{#1}\fi
\expandafter\ifx\csname bibfnamefont\endcsname\relax
  \def\bibfnamefont#1{#1}\fi
\expandafter\ifx\csname citenamefont\endcsname\relax
  \def\citenamefont#1{#1}\fi
\expandafter\ifx\csname url\endcsname\relax
  \def\url#1{\texttt{#1}}\fi
\expandafter\ifx\csname urlprefix\endcsname\relax\def\urlprefix{URL }\fi
\providecommand{\bibinfo}[2]{#2}
\providecommand{\eprint}[2][]{\url{#2}}

\bibitem{GS76}
\bibinfo{author}{\bibfnamefont{L.~C.} \bibnamefont{Gosule}} \bibnamefont{and}
  \bibinfo{author}{\bibfnamefont{J.~A.} \bibnamefont{Schellman}},
  \bibinfo{journal}{Nature} \textbf{\bibinfo{volume}{259}},
  \bibinfo{pages}{333} (\bibinfo{year}{1976}).

\bibitem{AB90}
\bibinfo{author}{\bibfnamefont{P.~G.} \bibnamefont{Arscott}} \bibnamefont{and}
\bibinfo{author}{\bibfnamefont{V.~A.} \bibnamefont{Bloomfield}},
  \bibinfo{journal}{Biopolymers} \textbf{\bibinfo{volume}{30}},
  \bibinfo{pages}{619} (\bibinfo{year}{1990}).

\bibitem{B91}
\bibinfo{author}{\bibfnamefont{V.~A.} \bibnamefont{Bloomfield}},
  \bibinfo{journal}{Biopolymers} \textbf{\bibinfo{volume}{31}},
  \bibinfo{pages}{1471} (\bibinfo{year}{1991}).

\bibitem{B96}
\bibinfo{author}{\bibfnamefont{V.~A.} \bibnamefont{Bloomfield}},
  \bibinfo{journal}{Curr.\ Opinion\ Struct.\ Biol.}
  \textbf{\bibinfo{volume}{6}}, \bibinfo{pages}{334} (\bibinfo{year}{1996}).

\bibitem{YYK98}
\bibinfo{author}{\bibfnamefont{Y.}~\bibnamefont{Yoshikawa}},
  \bibinfo{author}{\bibfnamefont{K.}~\bibnamefont{Yoshikawa}},
  \bibnamefont{and} \bibinfo{author}{\bibfnamefont{T.}~\bibnamefont{Kanbe}},
  \bibinfo{journal}{Langmuir} \textbf{\bibinfo{volume}{15}},
  \bibinfo{pages}{4085} (\bibinfo{year}{1999}).

\bibitem{CVH03}
\bibinfo{author}{\bibfnamefont{C.~C.} \bibnamefont{Conwell}},
  \bibinfo{author}{\bibfnamefont{I.~D.} \bibnamefont{Vilfan}},
  \bibnamefont{and} \bibinfo{author}{\bibfnamefont{N.~V.} \bibnamefont{Hud}},
  \bibinfo{journal}{Proc.\ Natl.\ Acad.\ Sci.} \textbf{\bibinfo{volume}{100}},
  \bibinfo{pages}{9296} (\bibinfo{year}{2003}).

\bibitem{HD01}
\bibinfo{author}{\bibfnamefont{N.~V.} \bibnamefont{Hud}} \bibnamefont{and}
  \bibinfo{author}{\bibfnamefont{K.~H.} \bibnamefont{Downing}},
  \bibinfo{journal}{Proc.\ Natl.\ Acad.\ Sci.} \textbf{\bibinfo{volume}{98}},
  \bibinfo{pages}{14925} (\bibinfo{year}{2001}).

\bibitem{DE86}
\bibinfo{author}{\bibfnamefont{M.}~\bibnamefont{Doi}} \bibnamefont{and}
  \bibinfo{author}{\bibfnamefont{S.~F.} \bibnamefont{Edwards}},
  \emph{\bibinfo{title}{The theory of polymer dynamics}}
  (\bibinfo{publisher}{Clarendon Press}, \bibinfo{address}{Oxford},
  \bibinfo{year}{1986}).

\bibitem{K04}
\bibinfo{author}{\bibfnamefont{H.}~\bibnamefont{Kleinert}},
  \emph{\bibinfo{title}{Path Integrals in Quantum Mechanics, Statistics, and
  Polymer Physics, and Financial Markets}} (\bibinfo{publisher}{World
  Scientific Publishing Company}, \bibinfo{year}{2004}).

\bibitem{KF72}
\bibinfo{author}{\bibfnamefont{K.~F.} \bibnamefont{Freed}},
  \bibinfo{journal}{Adv.\ Chem.\ Phys.} \textbf{\bibinfo{volume}{22}},
  \bibinfo{pages}{1} (\bibinfo{year}{1972}).

\bibitem{MPM04}
\bibinfo{author}{\bibfnamefont{A.}~\bibnamefont{Montesi}},
  \bibinfo{author}{\bibfnamefont{M.}~\bibnamefont{Pasquali}}, \bibnamefont{and}
  \bibinfo{author}{\bibfnamefont{F.~C.} \bibnamefont{MacKintosh}},
  \bibinfo{journal}{Phys.\ Rev.\ E} \textbf{\bibinfo{volume}{69}},
  \bibinfo{pages}{021916} (\bibinfo{year}{2004}).

\bibitem{CW04}
\bibinfo{author}{\bibfnamefont{I.~R.} \bibnamefont{Cooke}} \bibnamefont{and}
  \bibinfo{author}{\bibfnamefont{D.~R.~M.} \bibnamefont{Williams}},
  \bibinfo{journal}{Physica A} \textbf{\bibinfo{volume}{339}},
  \bibinfo{pages}{45} (\bibinfo{year}{2004}).

\bibitem{NSKY96}
\bibinfo{author}{\bibfnamefont{H.}~\bibnamefont{Noguchi}},
  \bibinfo{author}{\bibfnamefont{S.}~\bibnamefont{Saito}},
  \bibinfo{author}{\bibfnamefont{S.}~\bibnamefont{Kidoaki}}, \bibnamefont{and}
  \bibinfo{author}{\bibfnamefont{K.}~\bibnamefont{Yoshikawa}},
  \bibinfo{journal}{Chem.\ Phys.\ Lett.} \textbf{\bibinfo{volume}{261}},
  \bibinfo{pages}{527} (\bibinfo{year}{1996}).

\bibitem{NY98}
\bibinfo{author}{\bibfnamefont{H.}~\bibnamefont{Noguchi}} \bibnamefont{and}
  \bibinfo{author}{\bibfnamefont{K.}~\bibnamefont{Yoshikawa}},
  \bibinfo{journal}{J.\ Chem.\ Phys.} \textbf{\bibinfo{volume}{109}},
  \bibinfo{pages}{5070} (\bibinfo{year}{1998}).

\bibitem{KT99}
\bibinfo{author}{\bibfnamefont{Y.~A.} \bibnamefont{Kuznetsov}}
  \bibnamefont{and} \bibinfo{author}{\bibfnamefont{E.~G.}
  \bibnamefont{Timoshenko}}, \bibinfo{journal}{J.\ Chem.\ Phys.}
  \textbf{\bibinfo{volume}{111}}, \bibinfo{pages}{3744} (\bibinfo{year}{1999}).

\bibitem{KTD96}
\bibinfo{author}{\bibfnamefont{Y.~A.} \bibnamefont{Kuznetsov}},
  \bibinfo{author}{\bibfnamefont{E.~G.} \bibnamefont{Timoshenko}},
  \bibnamefont{and} \bibinfo{author}{\bibfnamefont{K.~A.}
  \bibnamefont{Dawson}}, \bibinfo{journal}{J.\ Chem.\ Phys.}
  \textbf{\bibinfo{volume}{105}}, \bibinfo{pages}{7116} (\bibinfo{year}{1996}).

\bibitem{IPB98}
\bibinfo{author}{\bibfnamefont{V.~A.} \bibnamefont{Ivanov}},
  \bibinfo{author}{\bibfnamefont{W.}~\bibnamefont{Paul}}, \bibnamefont{and}
  \bibinfo{author}{\bibfnamefont{K.}~\bibnamefont{Binder}},
  \bibinfo{journal}{J.\ Chem.\ Phys.} \textbf{\bibinfo{volume}{109}},
  \bibinfo{pages}{5659} (\bibinfo{year}{1998}).

\bibitem{GK81}
\bibinfo{author}{\bibfnamefont{A.~Y.} \bibnamefont{Grosberg}} \bibnamefont{and}
  \bibinfo{author}{\bibfnamefont{A.~R.} \bibnamefont{Khokhlov}},
  \bibinfo{journal}{Adv.\ Polym.\ Sci.} \textbf{\bibinfo{volume}{41}},
  \bibinfo{pages}{53} (\bibinfo{year}{1981}).

\bibitem{SIGPB03}
 
\bibinfo{author}{\bibfnamefont{M.~R.} \bibnamefont{Stukan}},
  \bibinfo{author}{\bibfnamefont{V.~A.} \bibnamefont{Ivanov}},
  \bibinfo{author}{\bibfnamefont{A.~Y.} \bibnamefont{Grosberg}},
  \bibinfo{author}{\bibfnamefont{W.}~\bibnamefont{Paul}}, \bibnamefont{and}
  \bibinfo{author}{\bibfnamefont{K.}~\bibnamefont{Binder}},
  \bibinfo{journal}{J.\ Chem.\ Phys.} \textbf{\bibinfo{volume}{118}},
  \bibinfo{pages}{3392} (\bibinfo{year}{2003}).

\bibitem{HDB95}
\bibinfo{author}{\bibfnamefont{N.~V.} \bibnamefont{Hud}},
  \bibinfo{author}{\bibfnamefont{K.~H.} \bibnamefont{Downing}},
  \bibnamefont{and} \bibinfo{author}{\bibfnamefont{R.}~\bibnamefont{Balhorn}},
  \bibinfo{journal}{Proc.\ Natl.\ Acad.\ Sci.} \textbf{\bibinfo{volume}{92}},
  \bibinfo{pages}{3581} (\bibinfo{year}{1995}).

\bibitem{UO95}
\bibinfo{author}{\bibfnamefont{J.}~\bibnamefont{Ubbink}} \bibnamefont{and}
  \bibinfo{author}{\bibfnamefont{T.}~\bibnamefont{Odijk}},
  \bibinfo{journal}{Biophys.\ J.} \textbf{\bibinfo{volume}{68}},
  \bibinfo{pages}{54} (\bibinfo{year}{1995}).

\bibitem{UO96}
\bibinfo{author}{\bibfnamefont{J.}~\bibnamefont{Ubbink}} \bibnamefont{and}
  \bibinfo{author}{\bibfnamefont{T.}~\bibnamefont{Odijk}},
  \bibinfo{journal}{Europhys.\ Lett.} \textbf{\bibinfo{volume}{33}},
  \bibinfo{pages}{353} (\bibinfo{year}{1996}).

\bibitem{SMW00}
\bibinfo{author}{\bibfnamefont{B.}~\bibnamefont{Schnurr}},
  \bibinfo{author}{\bibfnamefont{F.~C.} \bibnamefont{MacKintosh}},
  \bibnamefont{and} \bibinfo{author}{\bibfnamefont{D.~R.~M.}
  \bibnamefont{Williams}}, \bibinfo{journal}{Europhys.\ Lett.}
  \textbf{\bibinfo{volume}{51}}, \bibinfo{pages}{279} (\bibinfo{year}{2000}).

\bibitem{SGM02}
\bibinfo{author}{\bibfnamefont{B.}~\bibnamefont{Schnurr}},
  \bibinfo{author}{\bibfnamefont{F.}~\bibnamefont{Gittes}}, \bibnamefont{and}
  \bibinfo{author}{\bibfnamefont{F.~C.} \bibnamefont{MacKintosh}},
  \bibinfo{journal}{Phys.\ Rev.\ E} \textbf{\bibinfo{volume}{65}},
  \bibinfo{pages}{061904} (\bibinfo{year}{2002}).

\bibitem{PW00}
\bibinfo{author}{\bibfnamefont{G.~G.} \bibnamefont{Pereira}} \bibnamefont{and}
  \bibinfo{author}{\bibfnamefont{D.~R.~M.} \bibnamefont{Williams}},
  \bibinfo{journal}{Europhys.\ Lett.} \textbf{\bibinfo{volume}{50}},
  \bibinfo{pages}{559} (\bibinfo{year}{2000}).

\bibitem{MKPW05}
\bibinfo{author}{\bibfnamefont{I.~C.~B.} \bibnamefont{Miller}},
  \bibinfo{author}{\bibfnamefont{M.}~\bibnamefont{Keentok}},
  \bibinfo{author}{\bibfnamefont{G.~G.} \bibnamefont{Pereira}},
  \bibnamefont{and} \bibinfo{author}{\bibfnamefont{D.~R.~M.}
  \bibnamefont{Williams}}, \bibinfo{journal}{Phys.\ Rev.\ E}
  \textbf{\bibinfo{volume}{71}}, \bibinfo{pages}{031802}
  (\bibinfo{year}{2005}).

\bibitem{LGK78}
\bibinfo{author}{\bibfnamefont{I.~M.} \bibnamefont{Lifshitz}},
  \bibinfo{author}{\bibfnamefont{A.~Y.} \bibnamefont{Grosberg}},
  \bibnamefont{and} \bibinfo{author}{\bibfnamefont{A.~R.}
  \bibnamefont{Khokhlov}}, \bibinfo{journal}{Rev.\ Mod.\ Phys.}
  \textbf{\bibinfo{volume}{50}}, \bibinfo{pages}{683} (\bibinfo{year}{1978}).

\bibitem{KF84}
\bibinfo{author}{\bibfnamefont{A.~L.} \bibnamefont{Kholodenko}}
  \bibnamefont{and} \bibinfo{author}{\bibfnamefont{K.~F.} \bibnamefont{Freed}},
  \bibinfo{journal}{J.\ Phys.\ A:\ Math.\ Gen.} \textbf{\bibinfo{volume}{17}},
  \bibinfo{pages}{2703} (\bibinfo{year}{1984}).

\bibitem{DG79}
\bibinfo{author}{\bibfnamefont{P.~G.} \bibnamefont{de~Gennes}},
  \emph{\bibinfo{title}{Scaling Concepts in Polymer Physics}}
  (\bibinfo{publisher}{Cornell University Press}, \bibinfo{address}{New York},
  \bibinfo{year}{1979}).

\bibitem{GK94}
\bibinfo{author}{\bibfnamefont{A.~Y.} \bibnamefont{Grosberg}} \bibnamefont{and}
  \bibinfo{author}{\bibfnamefont{A.~R.} \bibnamefont{Khokhlov}},
  \emph{\bibinfo{title}{Statistical physics of macromolecules}}
  (\bibinfo{publisher}{American Institute of Physics}, \bibinfo{address}{New
  York}, \bibinfo{year}{1994}).

\bibitem{IK05}
\bibinfo{author}{\bibfnamefont{Y.}~\bibnamefont{Ishimoto}} \bibnamefont{and}
  \bibinfo{author}{\bibfnamefont{N.}~\bibnamefont{Kikuchi}},
  \bibinfo{journal}{J.\ Chem.\ Phys.} \textbf{\bibinfo{volume}{125}},
  \bibinfo{pages}{074905} (\bibinfo{year}{2006}).

\bibitem{C05}
\bibinfo{author}{\bibfnamefont{A.~G.} \bibnamefont{Cherstvy}},
  \bibinfo{journal}{J.\ Phys.:\ Condens.\ Matter} \textbf{\bibinfo{volume}{17}},
  \bibinfo{pages}{1363} (\bibinfo{year}{2005}).

\end{thebibliography}
\end{document}